\documentclass[prl,preprint,amsfonts,showpacs,showkeys]{revtex4}
\usepackage{graphicx}
\RequirePackage{times}
\RequirePackage{mathptm}

\begin{document}

\title{Communication cost of breaking the Bell barrier}

\author{Karl Svozil}
\email{svozil@tuwien.ac.at}
\homepage{http://tph.tuwien.ac.at/~svozil}
\affiliation{Institut f\"ur Theoretische Physik, University of Technology Vienna,
Wiedner Hauptstra\ss e 8-10/136, A-1040 Vienna, Austria}

\begin{abstract}
Correlations in an Einstein-Podolsky-Rosen-Bohm experiment can be made stronger than quantum correlations by allowing a single bit of classical communication between the two sides of the experiment.
\end{abstract}


\maketitle

From an operational point of view,
the nonlocal quantum correlations giving rise to
violations of Bell-type inequalities amount to the fact that
certain joint events at spacelike separated locations
occur with greater or smaller frequencies than can possibly be
expected from classical, local realistic models.
Two detectors at different locations register pairs
of particles or particle properties
more frequently or infrequently as can be explained by
the usual classical assumptions such as value definiteness.

With the rise of quantum algorithms and quantum information theory
\cite{nielsen-book},
the emphasis shifted to the communication cost and to the quantum communication complexity
related to those quantum correlations.
The question of the expense of obtaining quantum-type correlations
from classical systems was stimulated by quantum
\cite{BBCJPW}
and classical \cite{Brassard-Cleve-Tapp,cerf-gisin-massar-00,cerf-gisin-massar-pop-04,bru-gis-sca-04}
teleportation.
In a recent Letter~\cite{toner-bacon-03}, Toner and Bacon,
based on Refs. \cite{schatten-93,cerf-gisin-massar-00}, argue
that classical systems could mimic quantum systems by reproducing the
cosine law for correlation functions with the exchange of just one bit of classical information.

The formal coincidence of the quantum correlation function with classical
correlations augmented with the exchange of a single classical bit might indicate a deep structure
in quantum correlations.
One could, for instance, speculate that two-partite quantum systems may be capable of
conferring a single bit, a property which is reflected by the cosine form of the expectation function.
In what follows it will be argued that, while this may still be the case for the Toner-Bacon
protocol~\cite{toner-bacon-03}, in general the exchange of a single classical bit can give rise to
stronger than quantum correlations.

Since the systems discussed are entirely planar, whenever possible,
polar angles are used to represent the associated unit vectors.
The same symbols denote polar angles (without hat)
and the associated vectors (with hat).
Consider two correlated and spatially separated classical subsystems
sharing common
directions $ { \lambda }_i$, $i=1,\ldots$
which are chosen independently of each other and
 are distributed uniformly.
All parameters $ {\lambda }_i$ are assumed to be identical on each one of the two subsystems.
There are two measurement directions ${ a}$ and ${ b}$
of two
dichotomic observables with values ``-1'' and ``1''
at two spatially separated locations.
The measurement direction ${a}$ at ``Alice's location''
is unknown to an observer ``Bob'' measuring ${ b}$ and {\it vice versa}.
A two-particle correlation function $E(\theta )$
with $\theta =\vert {a} - { b}\vert $
is defined by averaging the product of the outcomes $O({ a})_i, O({ b} )_i\in {-1,1}$
in the $i$th experiment; i.e.,  $E(\theta )=(1/N)\sum_{i=1}^N O({ a})_i O({ b})_i$.


The following nonadaptive, memoryless protocols
could give rise to stronger-than-quantum correlations
by allowing the exchange of a single bit per experiment.
The protocols are similar to the one discussed by
Toner and Bacon~\cite{toner-bacon-03}, but require only a single share ${ \lambda }$,
and an additional direction ${ \Delta} (\delta )$, which is obtained by rotating
${\hat \lambda }$ clockwise around the origin by an angle $\delta$ which is a constant
shift for all experiments.
That is, ${\Delta} (\delta )=\lambda +\delta$.
Alice's observable is given by
$\alpha  = {\rm sgn}({\hat a} \cdot {\hat \lambda } )
={\rm sgn}\left[\cos ({a} - { \lambda } )\right]$.
The bit communicated by Alice is given by
$
c(\delta) =
{\rm sgn}({\hat a} \cdot {\hat \lambda } )
{\rm sgn}\left[{\hat a} \cdot {\hat \Delta} (\delta)\right]=
{\rm sgn}\left[\cos ({ a} - { \lambda } )\right]
{\rm sgn}\cos \left[{ a} - { \Delta} (\delta)\right]
$.
Bob's observable is  defined by
$\beta (\delta )=  {\rm sgn}[{\hat b} \cdot ({\hat \lambda } +c(\delta){\hat \Delta} (\delta))]$.
This protocol becomes Toner and Bacon's if $\delta$
is allowed to vary randomly, with uniform distribution.

\begin{figure}[htbp]
  \centering
\unitlength 1.00mm
\linethickness{0.5pt}
\begin{picture}(63.00,65.84)
\put(30.00,60.00){\line(1,0){1.61}}
\multiput(31.61,59.96)(0.80,-0.06){2}{\line(1,0){0.80}}
\multiput(33.22,59.83)(0.80,-0.11){2}{\line(1,0){0.80}}
\multiput(34.81,59.61)(0.53,-0.10){3}{\line(1,0){0.53}}
\multiput(36.39,59.31)(0.39,-0.10){4}{\line(1,0){0.39}}
\multiput(37.96,58.93)(0.39,-0.12){4}{\line(1,0){0.39}}
\multiput(39.50,58.46)(0.30,-0.11){5}{\line(1,0){0.30}}
\multiput(41.01,57.91)(0.25,-0.11){6}{\line(1,0){0.25}}
\multiput(42.50,57.27)(0.24,-0.12){6}{\line(1,0){0.24}}
\multiput(43.94,56.56)(0.20,-0.11){7}{\line(1,0){0.20}}
\multiput(45.35,55.78)(0.17,-0.11){8}{\line(1,0){0.17}}
\multiput(46.71,54.92)(0.16,-0.12){8}{\line(1,0){0.16}}
\multiput(48.02,53.98)(0.14,-0.11){9}{\line(1,0){0.14}}
\multiput(49.28,52.98)(0.13,-0.12){9}{\line(1,0){0.13}}
\multiput(50.49,51.91)(0.11,-0.11){10}{\line(1,0){0.11}}
\multiput(51.64,50.78)(0.11,-0.12){10}{\line(0,-1){0.12}}
\multiput(52.72,49.59)(0.11,-0.14){9}{\line(0,-1){0.14}}
\multiput(53.74,48.34)(0.12,-0.16){8}{\line(0,-1){0.16}}
\multiput(54.69,47.04)(0.11,-0.17){8}{\line(0,-1){0.17}}
\multiput(55.57,45.69)(0.12,-0.20){7}{\line(0,-1){0.20}}
\multiput(56.37,44.30)(0.10,-0.21){7}{\line(0,-1){0.21}}
\multiput(57.10,42.86)(0.11,-0.25){6}{\line(0,-1){0.25}}
\multiput(57.75,41.39)(0.11,-0.30){5}{\line(0,-1){0.30}}
\multiput(58.33,39.88)(0.10,-0.31){5}{\line(0,-1){0.31}}
\multiput(58.82,38.35)(0.10,-0.39){4}{\line(0,-1){0.39}}
\multiput(59.22,36.79)(0.11,-0.53){3}{\line(0,-1){0.53}}
\multiput(59.54,35.21)(0.12,-0.80){2}{\line(0,-1){0.80}}
\multiput(59.78,33.62)(0.08,-0.80){2}{\line(0,-1){0.80}}
\put(59.93,32.01){\line(0,-1){1.61}}
\put(60.00,30.40){\line(0,-1){1.61}}
\put(59.98,28.79){\line(0,-1){1.61}}
\multiput(59.87,27.18)(-0.10,-0.80){2}{\line(0,-1){0.80}}
\multiput(59.67,25.59)(-0.09,-0.53){3}{\line(0,-1){0.53}}
\multiput(59.39,24.00)(-0.09,-0.39){4}{\line(0,-1){0.39}}
\multiput(59.03,22.43)(-0.11,-0.39){4}{\line(0,-1){0.39}}
\multiput(58.58,20.88)(-0.11,-0.30){5}{\line(0,-1){0.30}}
\multiput(58.05,19.36)(-0.10,-0.25){6}{\line(0,-1){0.25}}
\multiput(57.44,17.87)(-0.12,-0.24){6}{\line(0,-1){0.24}}
\multiput(56.75,16.42)(-0.11,-0.20){7}{\line(0,-1){0.20}}
\multiput(55.98,15.00)(-0.11,-0.17){8}{\line(0,-1){0.17}}
\multiput(55.14,13.63)(-0.11,-0.17){8}{\line(0,-1){0.17}}
\multiput(54.22,12.30)(-0.11,-0.14){9}{\line(0,-1){0.14}}
\multiput(53.24,11.03)(-0.12,-0.14){9}{\line(0,-1){0.14}}
\multiput(52.19,9.81)(-0.11,-0.12){10}{\line(0,-1){0.12}}
\multiput(51.07,8.64)(-0.12,-0.11){10}{\line(-1,0){0.12}}
\multiput(49.89,7.54)(-0.14,-0.12){9}{\line(-1,0){0.14}}
\multiput(48.66,6.51)(-0.14,-0.11){9}{\line(-1,0){0.14}}
\multiput(47.37,5.54)(-0.17,-0.11){8}{\line(-1,0){0.17}}
\multiput(46.03,4.64)(-0.20,-0.12){7}{\line(-1,0){0.20}}
\multiput(44.65,3.82)(-0.20,-0.11){7}{\line(-1,0){0.20}}
\multiput(43.22,3.07)(-0.24,-0.11){6}{\line(-1,0){0.24}}
\multiput(41.76,2.40)(-0.30,-0.12){5}{\line(-1,0){0.30}}
\multiput(40.26,1.81)(-0.31,-0.10){5}{\line(-1,0){0.31}}
\multiput(38.73,1.30)(-0.39,-0.11){4}{\line(-1,0){0.39}}
\multiput(37.18,0.87)(-0.52,-0.11){3}{\line(-1,0){0.52}}
\multiput(35.61,0.53)(-0.53,-0.09){3}{\line(-1,0){0.53}}
\multiput(34.02,0.27)(-0.80,-0.09){2}{\line(-1,0){0.80}}
\put(32.41,0.10){\line(-1,0){1.61}}
\put(30.81,0.01){\line(-1,0){1.61}}
\put(29.19,0.01){\line(-1,0){1.61}}
\multiput(27.59,0.10)(-0.80,0.09){2}{\line(-1,0){0.80}}
\multiput(25.98,0.27)(-0.53,0.09){3}{\line(-1,0){0.53}}
\multiput(24.39,0.53)(-0.52,0.11){3}{\line(-1,0){0.52}}
\multiput(22.82,0.87)(-0.39,0.11){4}{\line(-1,0){0.39}}
\multiput(21.27,1.30)(-0.31,0.10){5}{\line(-1,0){0.31}}
\multiput(19.74,1.81)(-0.30,0.12){5}{\line(-1,0){0.30}}
\multiput(18.24,2.40)(-0.24,0.11){6}{\line(-1,0){0.24}}
\multiput(16.78,3.07)(-0.20,0.11){7}{\line(-1,0){0.20}}
\multiput(15.35,3.82)(-0.20,0.12){7}{\line(-1,0){0.20}}
\multiput(13.97,4.64)(-0.17,0.11){8}{\line(-1,0){0.17}}
\multiput(12.63,5.54)(-0.14,0.11){9}{\line(-1,0){0.14}}
\multiput(11.34,6.51)(-0.14,0.12){9}{\line(-1,0){0.14}}
\multiput(10.11,7.54)(-0.12,0.11){10}{\line(-1,0){0.12}}
\multiput(8.93,8.64)(-0.11,0.12){10}{\line(0,1){0.12}}
\multiput(7.81,9.81)(-0.12,0.14){9}{\line(0,1){0.14}}
\multiput(6.76,11.03)(-0.11,0.14){9}{\line(0,1){0.14}}
\multiput(5.78,12.30)(-0.11,0.17){8}{\line(0,1){0.17}}
\multiput(4.86,13.63)(-0.11,0.17){8}{\line(0,1){0.17}}
\multiput(4.02,15.00)(-0.11,0.20){7}{\line(0,1){0.20}}
\multiput(3.25,16.42)(-0.12,0.24){6}{\line(0,1){0.24}}
\multiput(2.56,17.87)(-0.10,0.25){6}{\line(0,1){0.25}}
\multiput(1.95,19.36)(-0.11,0.30){5}{\line(0,1){0.30}}
\multiput(1.42,20.88)(-0.11,0.39){4}{\line(0,1){0.39}}
\multiput(0.97,22.43)(-0.09,0.39){4}{\line(0,1){0.39}}
\multiput(0.61,24.00)(-0.09,0.53){3}{\line(0,1){0.53}}
\multiput(0.33,25.59)(-0.10,0.80){2}{\line(0,1){0.80}}
\put(0.13,27.18){\line(0,1){1.61}}
\put(0.02,28.79){\line(0,1){1.61}}
\put(0.00,30.40){\line(0,1){1.61}}
\multiput(0.07,32.01)(0.08,0.80){2}{\line(0,1){0.80}}
\multiput(0.22,33.62)(0.12,0.80){2}{\line(0,1){0.80}}
\multiput(0.46,35.21)(0.11,0.53){3}{\line(0,1){0.53}}
\multiput(0.78,36.79)(0.10,0.39){4}{\line(0,1){0.39}}
\multiput(1.18,38.35)(0.10,0.31){5}{\line(0,1){0.31}}
\multiput(1.67,39.88)(0.11,0.30){5}{\line(0,1){0.30}}
\multiput(2.25,41.39)(0.11,0.25){6}{\line(0,1){0.25}}
\multiput(2.90,42.86)(0.10,0.21){7}{\line(0,1){0.21}}
\multiput(3.63,44.30)(0.12,0.20){7}{\line(0,1){0.20}}
\multiput(4.43,45.69)(0.11,0.17){8}{\line(0,1){0.17}}
\multiput(5.31,47.04)(0.12,0.16){8}{\line(0,1){0.16}}
\multiput(6.26,48.34)(0.11,0.14){9}{\line(0,1){0.14}}
\multiput(7.28,49.59)(0.11,0.12){10}{\line(0,1){0.12}}
\multiput(8.36,50.78)(0.11,0.11){10}{\line(1,0){0.11}}
\multiput(9.51,51.91)(0.13,0.12){9}{\line(1,0){0.13}}
\multiput(10.72,52.98)(0.14,0.11){9}{\line(1,0){0.14}}
\multiput(11.98,53.98)(0.16,0.12){8}{\line(1,0){0.16}}
\multiput(13.29,54.92)(0.17,0.11){8}{\line(1,0){0.17}}
\multiput(14.65,55.78)(0.20,0.11){7}{\line(1,0){0.20}}
\multiput(16.06,56.56)(0.24,0.12){6}{\line(1,0){0.24}}
\multiput(17.50,57.27)(0.25,0.11){6}{\line(1,0){0.25}}
\multiput(18.99,57.91)(0.30,0.11){5}{\line(1,0){0.30}}
\multiput(20.50,58.46)(0.39,0.12){4}{\line(1,0){0.39}}
\multiput(22.04,58.93)(0.39,0.10){4}{\line(1,0){0.39}}
\multiput(23.61,59.31)(0.53,0.10){3}{\line(1,0){0.53}}
\multiput(25.19,59.61)(0.80,0.11){2}{\line(1,0){0.80}}
\multiput(26.78,59.83)(1.61,0.09){2}{\line(1,0){1.61}}
\put(30.00,30.00){\vector(2,1){26.33}}
\put(16.50,57.00){\line(1,-2){26.83}}
\put(30.00,65.84){\makebox(0,0)[cc]{${\hat b}$}}
\put(63.00,46.00){\makebox(0,0)[cc]{${\hat a}$}}
\put(46.00,51.83){\makebox(0,0)[cc]{$+1$}}
\put(7.67,43.01){\makebox(0,0)[cc]{$-1$}}
\put(52.83,17.17){\makebox(0,0)[cc]{$-1$}}
\put(11.83,10.67){\makebox(0,0)[cc]{$+1$}}
\put(30.00,45.00){\line(1,0){1.00}}
\put(31.00,44.97){\line(1,0){0.99}}
\multiput(31.99,44.87)(0.49,-0.08){2}{\line(1,0){0.49}}
\multiput(32.97,44.70)(0.49,-0.11){2}{\line(1,0){0.49}}
\multiput(33.94,44.47)(0.32,-0.10){3}{\line(1,0){0.32}}
\multiput(34.90,44.18)(0.31,-0.12){3}{\line(1,0){0.31}}
\multiput(35.83,43.82)(0.23,-0.10){4}{\line(1,0){0.23}}
\multiput(36.73,43.40)(0.22,-0.12){4}{\line(1,0){0.22}}
\multiput(37.61,42.93)(0.17,-0.11){5}{\line(1,0){0.17}}
\multiput(38.45,42.39)(0.16,-0.12){5}{\line(1,0){0.16}}
\multiput(39.25,41.80)(0.13,-0.11){6}{\line(1,0){0.13}}
\multiput(40.02,41.16)(0.12,-0.12){6}{\line(1,0){0.12}}
\multiput(40.74,40.47)(0.11,-0.12){6}{\line(0,-1){0.12}}
\multiput(41.41,39.74)(0.10,-0.13){6}{\line(0,-1){0.13}}
\multiput(42.03,38.96)(0.11,-0.16){5}{\line(0,-1){0.16}}
\multiput(42.60,38.14)(0.10,-0.17){5}{\line(0,-1){0.17}}
\multiput(43.11,37.28)(0.11,-0.22){4}{\line(0,-1){0.22}}
\multiput(43.57,36.40)(0.10,-0.23){4}{\line(0,-1){0.23}}
\multiput(43.96,35.48)(0.11,-0.31){3}{\line(0,-1){0.31}}
\multiput(44.30,34.54)(0.09,-0.32){3}{\line(0,-1){0.32}}
\multiput(44.57,33.58)(0.10,-0.49){2}{\line(0,-1){0.49}}
\multiput(44.77,32.60)(0.07,-0.49){2}{\line(0,-1){0.49}}
\put(44.91,31.62){\line(0,-1){0.99}}
\put(44.99,30.62){\line(0,-1){1.00}}
\put(45.00,29.63){\line(0,-1){1.00}}
\multiput(44.94,28.63)(-0.06,-0.49){2}{\line(0,-1){0.49}}
\multiput(44.81,27.64)(-0.09,-0.49){2}{\line(0,-1){0.49}}
\multiput(44.62,26.66)(-0.08,-0.32){3}{\line(0,-1){0.32}}
\multiput(44.37,25.70)(-0.11,-0.32){3}{\line(0,-1){0.32}}
\multiput(44.05,24.75)(-0.09,-0.23){4}{\line(0,-1){0.23}}
\multiput(43.67,23.83)(-0.11,-0.22){4}{\line(0,-1){0.22}}
\multiput(43.23,22.94)(-0.10,-0.17){5}{\line(0,-1){0.17}}
\multiput(42.73,22.07)(-0.11,-0.17){5}{\line(0,-1){0.17}}
\multiput(42.18,21.24)(-0.10,-0.13){6}{\line(0,-1){0.13}}
\multiput(41.57,20.45)(-0.11,-0.12){6}{\line(0,-1){0.12}}
\multiput(40.91,19.71)(-0.12,-0.12){6}{\line(-1,0){0.12}}
\multiput(40.20,19.00)(-0.13,-0.11){6}{\line(-1,0){0.13}}
\multiput(39.45,18.35)(-0.13,-0.10){6}{\line(-1,0){0.13}}
\multiput(38.65,17.75)(-0.17,-0.11){5}{\line(-1,0){0.17}}
\multiput(37.82,17.20)(-0.17,-0.10){5}{\line(-1,0){0.17}}
\multiput(36.95,16.71)(-0.22,-0.11){4}{\line(-1,0){0.22}}
\multiput(36.06,16.28)(-0.23,-0.09){4}{\line(-1,0){0.23}}
\multiput(35.13,15.90)(-0.32,-0.10){3}{\line(-1,0){0.32}}
\multiput(34.18,15.59)(-0.32,-0.08){3}{\line(-1,0){0.32}}
\multiput(33.22,15.35)(-0.49,-0.09){2}{\line(-1,0){0.49}}
\put(32.24,15.17){\line(-1,0){0.99}}
\put(31.25,15.05){\line(-1,0){1.00}}
\put(30.25,15.00){\line(-1,0){1.00}}
\put(29.25,15.02){\line(-1,0){0.99}}
\multiput(28.26,15.10)(-0.49,0.07){2}{\line(-1,0){0.49}}
\multiput(27.27,15.25)(-0.49,0.11){2}{\line(-1,0){0.49}}
\multiput(26.30,15.46)(-0.32,0.09){3}{\line(-1,0){0.32}}
\multiput(25.34,15.74)(-0.31,0.11){3}{\line(-1,0){0.31}}
\multiput(24.40,16.08)(-0.23,0.10){4}{\line(-1,0){0.23}}
\multiput(23.49,16.49)(-0.22,0.12){4}{\line(-1,0){0.22}}
\multiput(22.61,16.95)(-0.17,0.10){5}{\line(-1,0){0.17}}
\multiput(21.76,17.47)(-0.16,0.12){5}{\line(-1,0){0.16}}
\multiput(20.94,18.04)(-0.13,0.10){6}{\line(-1,0){0.13}}
\multiput(20.17,18.67)(-0.12,0.11){6}{\line(-1,0){0.12}}
\multiput(19.44,19.35)(-0.11,0.12){6}{\line(0,1){0.12}}
\multiput(18.75,20.07)(-0.11,0.13){6}{\line(0,1){0.13}}
\multiput(18.12,20.84)(-0.12,0.16){5}{\line(0,1){0.16}}
\multiput(17.54,21.65)(-0.11,0.17){5}{\line(0,1){0.17}}
\multiput(17.01,22.50)(-0.12,0.22){4}{\line(0,1){0.22}}
\multiput(16.54,23.38)(-0.10,0.23){4}{\line(0,1){0.23}}
\multiput(16.13,24.29)(-0.12,0.31){3}{\line(0,1){0.31}}
\multiput(15.78,25.22)(-0.10,0.32){3}{\line(0,1){0.32}}
\multiput(15.50,26.18)(-0.11,0.49){2}{\line(0,1){0.49}}
\multiput(15.27,27.15)(-0.08,0.49){2}{\line(0,1){0.49}}
\put(15.12,28.13){\line(0,1){0.99}}
\put(15.03,29.13){\line(0,1){1.00}}
\put(15.00,30.12){\line(0,1){1.00}}
\put(15.04,31.12){\line(0,1){0.99}}
\multiput(15.15,32.11)(0.09,0.49){2}{\line(0,1){0.49}}
\multiput(15.32,33.09)(0.12,0.48){2}{\line(0,1){0.48}}
\multiput(15.56,34.06)(0.10,0.32){3}{\line(0,1){0.32}}
\multiput(15.86,35.01)(0.09,0.23){4}{\line(0,1){0.23}}
\multiput(16.23,35.94)(0.11,0.23){4}{\line(0,1){0.23}}
\multiput(16.65,36.84)(0.10,0.17){5}{\line(0,1){0.17}}
\multiput(17.14,37.71)(0.11,0.17){5}{\line(0,1){0.17}}
\multiput(17.68,38.55)(0.12,0.16){5}{\line(0,1){0.16}}
\multiput(18.27,39.35)(0.11,0.13){6}{\line(0,1){0.13}}
\multiput(18.92,40.11)(0.12,0.12){6}{\line(0,1){0.12}}
\multiput(19.62,40.82)(0.12,0.11){6}{\line(1,0){0.12}}
\multiput(20.36,41.49)(0.13,0.10){6}{\line(1,0){0.13}}
\multiput(21.14,42.11)(0.16,0.11){5}{\line(1,0){0.16}}
\multiput(21.97,42.67)(0.17,0.10){5}{\line(1,0){0.17}}
\multiput(22.83,43.17)(0.22,0.11){4}{\line(1,0){0.22}}
\multiput(23.72,43.62)(0.23,0.10){4}{\line(1,0){0.23}}
\multiput(24.64,44.01)(0.31,0.11){3}{\line(1,0){0.31}}
\multiput(25.58,44.33)(0.32,0.09){3}{\line(1,0){0.32}}
\multiput(26.54,44.60)(0.49,0.10){2}{\line(1,0){0.49}}
\multiput(27.52,44.79)(0.49,0.07){2}{\line(1,0){0.49}}
\put(28.51,44.93){\line(1,0){1.49}}
\put(30.00,30.00){\vector(4,-1){14.33}}
\put(30.00,30.00){\vector(-1,-4){3.58}}
\put(30.00,30.00){\line(1,4){3.63}}
\put(30.00,30.00){\line(-4,1){14.50}}
\put(42.33,33.67){\makebox(0,0)[cc]{$-$}}
\put(37.33,20.50){\makebox(0,0)[cc]{$+$}}
\put(19.17,24.17){\makebox(0,0)[cc]{$-$}}
\put(22.67,40.17){\makebox(0,0)[cc]{$+$}}
\put(47.00,25.67){\makebox(0,0)[cc]{${\hat \lambda}$}}
\put(25.67,12.00){\makebox(0,0)[cc]{${\hat \lambda}^\perp$}}
\put(48.00,44.50){\makebox(0,0)[cc]{${\hat \lambda}-{\hat \lambda}^\perp$}}
\put(30.00,30.02){\vector(3,2){18.60}}
\put(44.73,26.36){\vector(1,4){3.95}}
\put(30.00,30.00){\vector(0,1){30.00}}
\put(0.00,30.00){\line(1,0){60.00}}
\end{picture}
  \caption{Demonstration of the shift mechanism.
Concentric circles represent the measurement directions
${\hat a}$ and ${\hat  b}$ (outer circle), as well as parameter directions
${\hat  \lambda }$ and ${\hat  \lambda }^\perp$ (inner circle)
and their associated projective sign regions.
The four measurement regions spanned by
${\hat a}$ and ${\hat b}$ are indicated by ``$\pm 1$,'' respectively.
Positive and negative octants spanned by
${\hat  \lambda }$ and ${\hat  \lambda }^\perp$ are indiated in the inner circle by ``$\pm$,'' respectively.
In this configuration, the shift ${\hat \lambda } \longrightarrow  {\hat \lambda }-{\hat \lambda }^\perp$
pushes ${\hat \lambda }$ into a positive region.}
\label{2004-brainteaser-f0}
\end{figure}
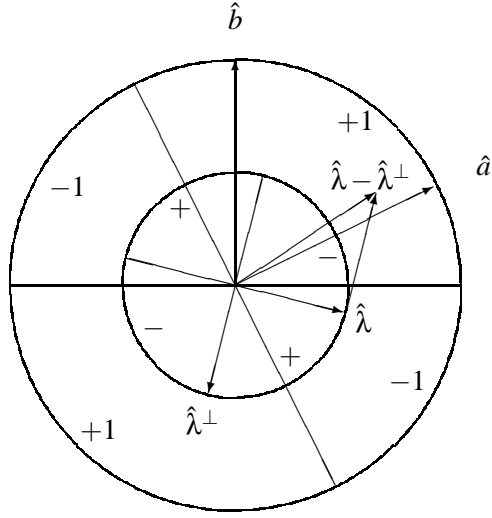

The strongest correlations are obtained for $\delta =\pi /2$; i.e., in the case
where the two directions associated with ${ \lambda }$ and ${ \Delta} ={ \lambda }+ \pi /2$
are orthogonal
and the information obtained by $c (\pi /2)$ is about the location of ${ a}$
within two opposite quadrants.
Let $H(x)$ stand for the Heaviside step function of $x$.
The effective shift in the parameter direction
${\hat \lambda }\longrightarrow {\hat \lambda } \pm {\hat \lambda }^\perp$
yields a correlation function of the form
\begin{equation}
E\left(\theta ,\delta = {\pi \over 2}\right)
=
H\left(\theta - {3 \pi\over 4}\right)
 - H\left({\pi \over 4} - \theta \right)  -
  2\left(1 - {2\over \pi}\theta \right)
H\left(\theta - {\pi\over 4}\right) H\left({3 \pi\over 4} - \theta \right)
.
\label{e-2004-brainteaser-2a}
\end{equation}
To obtain a better understanding for the shift mechanism, in
Fig.~\ref{2004-brainteaser-f0} a configuration is drawn which, without the shift
${\hat \lambda } \longrightarrow {\hat \lambda }-{\hat \lambda }^\perp$,
${\rm sgn}({\hat b} \cdot {\hat \lambda })={\rm sgn} \cos({b} - { \lambda })$
would have contributed the factor $-1$.
The shift
results in
a positive contribution ${\rm sgn}[{\hat b} \cdot ({\hat \lambda }-{\hat \lambda }^\perp )]$
to the expectation value.
This shift mechanism always yields the strongest correlations $\pm 1$ as long as
the angle $\theta$ does not lie between $\pi /4$ and $3 \pi /4$.

\begin{figure}[htbp]
  \centering
  \includegraphics[width=80mm]{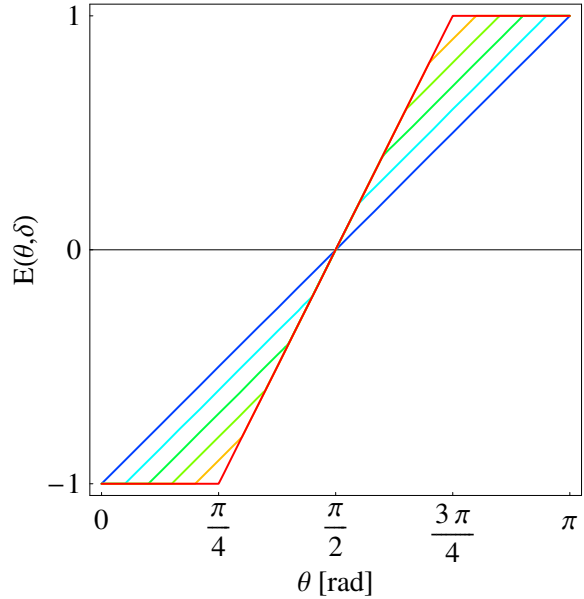}
  \caption{Classical and stronger-than-quantum correlation functions
obtained through the exchange of a single
bit in the memoryless regime for values of $\delta \in \{0,
{\pi \over 10},
{\pi \over 5},
{3\pi \over 10},
{2\pi \over 5},
{\pi \over 2}\}$ between $0$ (straigt line) and $\pi /2$ [cf. Eq~(\ref{e-2004-brainteaser-2a})].}
  \label{2004-brainteaser-f2}
\end{figure}
\begin{figure}[htbp]
  \centering
  \includegraphics[width=90mm]{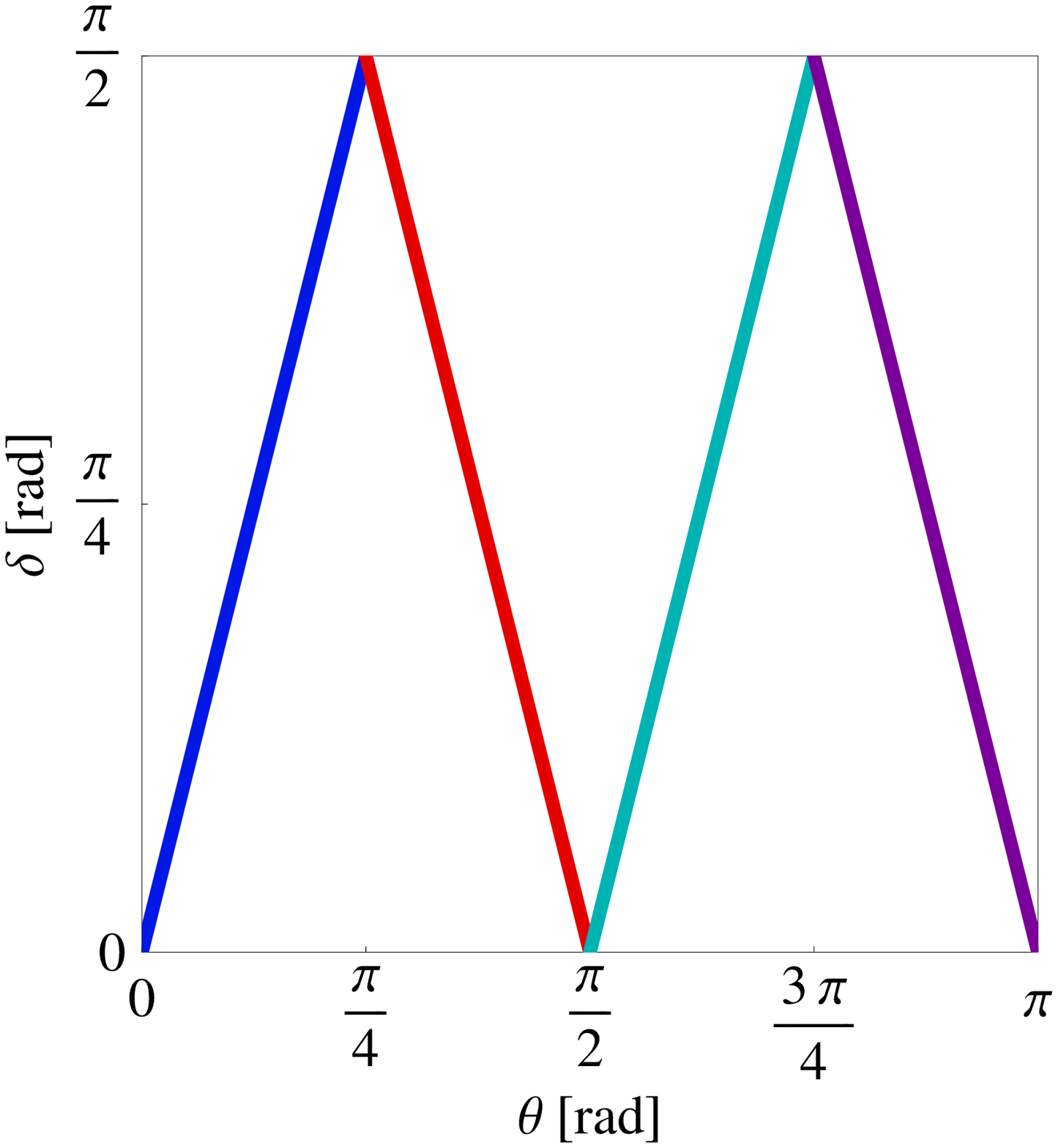}
  \caption{Domains of the correlation function $E(\theta ,\delta )$.
Consecutive sections from left to right cover the domains
           of Eq.~(\ref{e-2004-brainteaser-2}) from bottom to top:
(1) $0\le \theta \le {\delta \over 2}$,
(2) ${\delta \over 2} < \theta \le {1 \over 2}(\pi - \delta)$,
(3) ${1 \over 2}(\pi - \delta) < \theta \le {1 \over 2}(\pi   + \delta )$,
(4) ${1 \over 2}(\pi   + \delta )  < \theta \le \pi - {\delta \over 2}$,
(5) $\pi - {\delta \over 2} < \theta \le \pi$.
}
  \label{2004-brainteaser-f2b}
\end{figure}

For general $0\le \delta \le \pi /2$, Fig.~\ref{2004-brainteaser-f2} depicts numerical evaluations
which  fit the correlation function
\begin{equation}
E(\theta ,\delta )= \left\{
\begin{array}{ll}
-1 & \text{ for } \;\; 0\le \theta \le {\delta \over 2} ,  \\
-1 +{2\over \pi}(\theta -{\delta \over 2} )&\text{ for } \;\; {\delta \over 2} < \theta \le {1 \over 2}(\pi - \delta) ,   \\
-2(1-{2 \over \pi } \theta ) &\text{ for } \; \; {1 \over 2}(\pi - \delta) < \theta \le {1 \over 2}(\pi   + \delta ) , \\
1+ {2\over \pi }(\theta-\pi +{\delta \over 2} ) &\text{ for } \;\; {1 \over 2}(\pi   + \delta )  < \theta \le \pi - {\delta \over 2} , \\
1 & \text{ for } \;\; \pi - {\delta \over 2} < \theta \le \pi .
\end{array}        \right.
\label{e-2004-brainteaser-2}
\end{equation}
Its domains are depicted in Fig.~\ref{2004-brainteaser-f2b}.
For all nonzero $\delta$, $E(\theta ,\delta )$ correlates stronger than quantized systems
for some values of $\theta$.
With $\delta =\pi /2$, the Clauser-Horne-Shimony-Holt (CHSH) inequality
$
\vert
E({ a} ,{ b} )+
E({ a} ,{ b} ' )+
E({ a}' ,{ b} )-
E({a} ',{ b} ')
\vert
\le 2
$
for ${a} =\pi /2 $, ${a}'=0$, ${b} =\pi /4$, ${b}'= 3\pi /4$
is violated by the maximal algebraic value of 4 \cite{popescu-97b}, a larger value
than the Tsirelson bound for quantum violations $2\sqrt{2}$.
This is due to the fact that, phenomenologically, the strategy allows for
certain joint events to occur with greater or smaller frequencies as can be
expected from quantum entangled state measurements.
For $\delta =0$, the classical linear correlation function
$E(\theta )=  2\theta /\pi -1$ is recovered, as can be expected.

The average over all $0\le \delta \le \pi /2$
yields a very similar, but not identical behaviour
as the   quantum cosine correlation function.
More precisely,
assume two independent,
uniformly distributed shares ${\lambda}_1$ and ${\lambda}_2$, and
a single communicated bit $H\left[c({\hat  \lambda}_1,{\hat  \lambda}_2)\right]$ with
$c({\hat  \lambda}_1,{\hat  \lambda}_2)=
{\rm sgn}  ({\hat a} \cdot {\hat \lambda}_1){\rm sgn} ({\hat a} \cdot {\hat \lambda}_2)
= {\rm sgn} \cos ({ a} - { \lambda }_1 ){\rm sgn} \cos ({ a} - { \lambda }_2 )$
per measurement of
$
{\rm sgn} ({\hat a}\cdot {\hat \lambda}_1)
{\rm sgn}
[{\hat b}\cdot ({\hat \lambda}_2-c{\hat \lambda}_1)]
$.
The associated correlation function is
\begin{equation}
\begin{array}{cll}
&&E(\theta ) =
{1\over (2 \pi)^2}
\int d {\hat\lambda_1}  d {\hat\lambda_2}  \;
{\rm sgn}
({\hat a}\cdot {\hat \lambda}_1)
{\rm sgn}
[{\hat b}\cdot ({\hat \lambda}_2-c{\hat \lambda}_1)] \\
&&\qquad \qquad \qquad \qquad  =
{1\over 2 \pi^2}
\int d {\hat\lambda_1}  \;
{\rm sgn}
({\hat a}\cdot {\hat \lambda}_1)
\int d {\hat\lambda_2}    \;
{\rm sgn}
[{\hat b}\cdot ({\hat \lambda}_2-{\hat \lambda}_1)].
\end{array}
\label{e-2004-brainteaser-3}
\end{equation}

The elemination of $c$ at the cost of the prefactor $2$
on the right hand side of Eq.~(\ref{e-2004-brainteaser-3}) is achieved by
using the symmetries of the outcomes under the exchange
${\hat \lambda }_1 \longleftrightarrow -{\hat  \lambda }_1$
and
${\hat  \lambda }_2 \longleftrightarrow -{\hat  \lambda }_2$
as outlined in Ref.~\cite{toner-bacon-03}.
Note that, although the correlation function $E(\theta )$ is
nonlocal
(Bob's output depends on $c$, which contains ${ a}$),
after some recasting, despite the prefactor of 2 which accounts for nonlocality,
it appears to be perfectly local,
since it is the product of two sign functions
containing merely ${ a}$ and  (separately) ${b}$, respectively.
(The two parameters $ {\lambda }_1, {\lambda }_2$ are common shares.)

The ${\lambda}_2$ integration can be performed by
arranging
${\hat b}$ along the positive $y$-axis $(0,1)$ and by the parameterization
${\hat \lambda}_1=(\sin t ,\cos t )$
and
${\hat \lambda}_2=(\sin \tau ,\cos \tau )$.
The positive contributions amount to $A_+=2\int_0^t d\tau =2t$;
thus the negative contributions are $A_-= 2\pi -A_+$, and the entire integral
is $(A_+-A_-)/(2\pi )=2t/\pi -1=2\cos^{-1}({\hat b}\cdot {\hat \lambda}_1)/\pi -1$.

\begin{figure}[htbp]
  \centering
 \includegraphics[width=80mm]{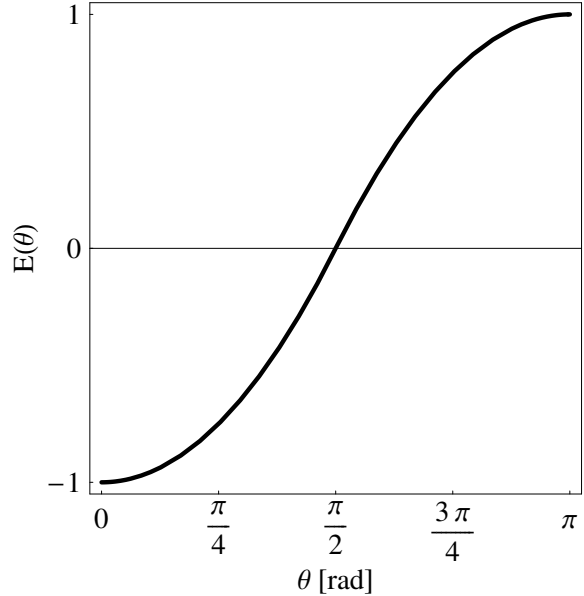}
  \caption{The correlation $E(\theta )$ of Eq.~(\ref{e-2004-brainteaser-4}) as a function of
$\theta =\vert{ a}- { b}\vert$ for the memoryless exchange of a single
bit per experiment in the planar configuration.
Note that, although the shape resembles the quantum cosine law,
the function is piecewise quadratic.}
  \label{2004-brainteaser-f3}
\end{figure}

For the ${\lambda}_1$ integration,
${\hat a}$ is arranged along the positive $y$-axis $(0,1)$,
${\hat b}$ along $(\sin r,\cos r)$, and
${\hat \lambda}_1$ along $(\sin \tau ,\cos \tau )$.
Since $\cos^{-1} (\cos (x))=\vert x\vert$ for $-1\le x\le 1$,
one obtains for the correlation value
$(4/\pi^2)\int_0^\pi d \tau \;{\rm sgn}(\cos \tau ) \vert \tau -r\vert$.
After evaluating all cases, the ${\hat \lambda}_1$ integration,
for $0\le \theta =\vert{ a}- { b}\vert \le \pi $, yields
\begin{equation}
E(\theta ) =
{4\over \pi^2}
\left[
\left(\theta ^2 -{\pi^2 \over 4}\right)
-2
H\left(\theta -{\pi \over 2}\right)
\left(\theta -{\pi \over 2} \right)^2
\right],
\label{e-2004-brainteaser-4}
\end{equation}
which is plotted in Fig.~\ref{2004-brainteaser-f3}.
An alternative derivation of Eq.~(\ref{e-2004-brainteaser-4}) is via
$(2/\pi )\int_0^{\pi /2} d\delta \; E(\theta,\delta )$ with $E(\theta,\delta )$
from Eq.~(\ref{e-2004-brainteaser-2}).
The difference to the cosine law obtained by Toner and Bacon~\cite{toner-bacon-03}
is due to the reduced dimensionality of the problem.


Let us briefly sketch a protocol requiring memory which,
by the exchange of more than one bit,
could give rise to maximal violations \cite{svozil-krenn}
of the CHSH
inequality.
The protocol  is based on locating and communicating information about
Alice's measurement direction
to Bob, who then rotates his subsystem
(or alternatively his measurement direction)
so as to obtain the desired correlation function.

When compared to the protocol discussed by Toner and Bacon \cite{toner-bacon-03}
or to the memoryless protocoll introduced above,
the adaptive protocols share some similarities.
Both exchange very similar information,
as can for instance be seen by a comparison between $c_2$ above
and the bit
$c= {\rm sgn} \cos ({ a} - { \lambda }_1 ){\rm sgn} \cos ({ a} - { \lambda }_2 )$
exchanged in the  Toner and Bacon protocol.
After the exchange of just a few bits, the adaptive protocols
appear to be more efficient
from the communication complexity point of view.
However, these strategies  require memory.
Operationally, this presents no problem for Alice and Bob,
but if one insist on nonadaptive, single particle strategies,
these protocols must be excluded.

In summary we have found that, as long as adaptive protocols
requiring memory at the receiver side are allowed,
the CHSH inequality can be violated maximally.
Furthermore, we have presented a type of memoryless, nonadaptive protocol
giving rise to stronger-than-quantum correlations which
yields the maximal algebraic violation $4$ of the CHSH inequality,
as compared to the quantum Tsirelson bound $2\sqrt{2}$.
The question  about the effect of an information exchange between entangled quantum subsystems still remains open.

The author acknowledges Akbar Fahmi for pointing out Ref.~\cite{popescu-97b}.


\end{document}